\documentclass[prd,aps,eqsecnum,showpacs,nofootinbib]{revtex4}

\usepackage{amsmath,amssymb}
\usepackage{bm}
\usepackage[dvips]{graphicx,color}

\begin{document}

\vglue 2cm

\title{REMARKS ON THE MYERS-PERRY AND EINSTEIN GAUSS-BONNET ROTATING
SOLUTIONS\footnote{It is a pleasure to dedicate this paper to Mario Castagnino
on the occasion of his ``Festschrift".}}
\author{ANDRES ANABALON}
\affiliation{Departamento de Ciencias, Facultad de Artes Liberales, Facultad de
Ingenier\'{\i}a y Ciencias, Universidad Adolfo Ib\'{a}\~{n}ez, Vi\~{n}a del
Mar, Chile.}
\author{NATHALIE DERUELLE}
\affiliation{APC, UMR 7164 du CNRS, Universit\'e Paris 7, 75205 Paris Cedex13, France}
\author{DAVID TEMPO$^{1}$ and RICARDO TRONCOSO$^{1,2}$}
\affiliation{$^{1}$Centro de Estudios Cient\'{\i}ficos (CECS), Casilla 1469, Valdivia, Chile, }
\affiliation{$^{2}$Centro de Ingenier\'{\i}a de la Innovaci\'{o}n del CECS (CIN), Valdivia, Chile.}
\date{\today}

\begin{abstract}
The Kerr-type solutions of the five-dimensional Einstein and
Einstein-Gauss-Bonnet equations look pretty similar when written in
Kerr-Schild form. However the Myers-Perry spacetime is circular whereas the
rotating solution of the Einstein-Gauss-Bonnet theory is not. We explore some
consequences of this difference in particular regarding the (non) existence of
Boyer-Lindquist-type coordinates and the extension of the manifold.

\end{abstract}

\maketitle

\section{Introduction}

Some interesting features of Kerr-Schild spacetimes with flat seed metrics in
four and five dimensions can be extracted from simple and purely geometrical
considerations. In what follows, the properties of spheroidal coordinates in
Euclidean flat space are reviewed, and in section 3 we unveil some nontrivial
aspects of the conditions for staticity and circularity of Kerr-Schild-type
metrics. The last section applies the previous analysis to exact rotating
solutions of the Einstein, and the Einstein-Gauss-Bonnet theories in vacuum.

\section{Euclidean spaces in spheroidal coordinates}

We compare here spheroidal coordinates in three \textsl{vs} four dimensional
Euclidean spaces. As we shall see, they cover the whole of $\mathbb{R}^{3}$
but not the whole of $\mathbb{R}^{4}$. \newline

Let us start with ordinary three-dimensional Euclidean space $\mathbb{R}^{3}$
in cartesian coordinates ($X,Y,Z$) and introduce new coordinates
($r,\vartheta,\varphi$) defined as
\begin{equation}
X=\sqrt{r^{2}+a^{2}}\sin\vartheta\cos\varphi\quad,\quad Y=\sqrt{r^{2}+a^{2}%
}\sin\vartheta\sin\varphi\quad,\quad Z=r\,\cos\vartheta\label{eqn1}%
\end{equation}
where $a$ is a parameter. Surfaces of constant $r>0$ are the spheroids
${\frac{X^{2}+Y^{2}}{r^{2}+a^{2}}}+{\frac{Z^{2}}{r^{2}}}=1$. Points on each
spheroid are defined by the angle $\varphi\in\lbrack0,2\pi\lbrack$ such that
$\tan\varphi={\frac{Y}{X}}$ together with the angle $\vartheta\in\lbrack
0,\pi]$ such that $\sin\vartheta=\sqrt{{\frac{X^{2}+Y^{2}}{r^{2}+a^{2}}}}$
(with $0<\vartheta<{\frac{\pi}{2}}$ if $Z>0$ and ${\frac{\pi}{2}}%
<\vartheta<\pi$ if $Z<0$). In the equatorial plane $Z=0$, points outside the
circle of radius $a$ are represented by $\vartheta={\frac{\pi}{2}}$, and
points within the circle by $r=0$. Thus the whole of $\mathbb{R}^{3}$ is
covered once by the coordinates ($r,\vartheta,\varphi$), apart from the origin
$X=Y=Z=0$.

In these spheroidal coordinates the line element reads
\begin{equation}
ds_{(3)}^{2}={\frac{\varrho^{2}}{r^{2}+a^{2}}}dr^{2}+\varrho^{2}d\vartheta
^{2}+(r^{2}+a^{2})\sin^{2}\vartheta\,d\varphi^{2}\quad\text{with}\quad
\varrho^{2}\equiv r^{2}+a^{2}\cos^{2}\vartheta\,. \label{eqn2}%
\end{equation}
The determinant of the metric is $\varrho^{4}\sin^{2}\vartheta$. It vanishes
on the $Z$ axis ($\vartheta=0$ or $\pi$), and also where $\varrho^{2}=0$, that
is, where $r=0$ and $\vartheta={\frac{\pi}{2}}$, which is the circle of radius
$a$ in the equatorial plane $Z=0$.\newline

Let us now turn to $\mathbb{R}^{4}$ in cartesian coordinates ($X,Y,Z,W$) and
introduce the system ($r,\vartheta,\varphi,\psi$) defined as
\begin{equation}%
\begin{array}
[c]{rcl}%
X & = & \sqrt{r^{2}+a^{2}}\sin\vartheta\cos\varphi\quad,\quad Y=\sqrt
{r^{2}+a^{2}}\sin\vartheta\sin\varphi\,,\\[8pt]%
Z & = & \sqrt{r^{2}+b^{2}}\,\cos\vartheta\cos\psi\quad,\quad W=\sqrt
{r^{2}+b^{2}}\,\cos\vartheta\sin\psi\,,
\end{array}
\label{eqn3}%
\end{equation}
where the parameters $a$ and $b$ can be chosen such that $a\geq b\geq0$. If we
impose as usual $r\in\lbrack0,\infty]$, $\vartheta\in\lbrack0,{\frac{\pi}{2}%
}]$, $\varphi\in\lbrack0,2\pi\lbrack$, $\psi\in\lbrack0,2\pi\lbrack$ (see
\textsl{e.g.} Ref.~\cite{Gibbons04}), then (when $b\neq0$) all points outside
the 3-surface defined by ${\frac{X^{2}+Y^{2}}{a^{2}}}+{\frac{Z^{2}+W^{2}%
}{b^{2}}}=1$ are represented once, but points inside this 3-spheroid, in
particular the origin, are \textsl{not} represented.

In these spheroidal coordinates the line element reads
\begin{equation}%
\begin{array}
[c]{rcl}%
ds_{(4)}^{2} & = & \Delta_{r}\,dr^{2}+\varrho^{2}d\vartheta^{2}+(r^{2}%
+a^{2})\sin^{2}\vartheta\,d\varphi^{2}+(r^{2}+b^{2})\cos^{2}\vartheta
\,d\psi^{2}\\[8pt]%
\quad & \text{with} & \quad\varrho^{2}\equiv r^{2}+a^{2}\cos^{2}%
\vartheta+b^{2}\sin^{2}\vartheta\quad\text{and}\quad\Delta_{r}={\frac
{r^{2}\varrho^{2}}{(r^{2}+a^{2})(r^{2}+b^{2})}}\,.
\end{array}
\label{eqn4}%
\end{equation}
Note that $\varrho^{2}$ never vanishes (unless $b=0$). The determinant is
$r^{2}\varrho^{4}\sin^{2}\vartheta\cos^{2}\vartheta$. It vanishes on the two
equatorial planes, $X=Y=0$ ($\vartheta=0$) and $Z=W=0$ ($\vartheta={\frac{\pi
}{2}}$), and also where $r=0$. This last coordinate singularity is due to the
fact that the coordinates ($r,\vartheta,\varphi,\psi$) do not cover the whole
of $\mathbb{R}^{4}$. As noted by Myers and Perry \cite{Myers86} (see also
Ref.~\cite{Gibbons09}), this last singularity is removed and the whole of
$\mathbb{R}^{4}$ is covered by simply changing $r$ into $x=r^{2}$ and allowing
$x\in\lbrack-b^{2},\infty]$. The coordinate transformation $r\rightarrow x$
becomes complex when $x<0$ but the line element remains real. The origin is
then represented by $x=-b^{2}$, $\vartheta=0$. As for $\varrho^{2}$ it can
then vanish at $x=-b^{2}$ and $\vartheta={\frac{\pi}{2}}$, which is the circle
of radius $\sqrt{a^{2}-b^{2}}$ in the equatorial plane $Z=W=0$.

\section{Staticity and circularity of $4D$ \textsl{vs} $5D$ Kerr-Schild
spacetimes}

In this section we shall first find a geometrical curiosity, that is that, in
$5D$, Kerr-Schild spacetimes can be static for equal but non vanishing
parameters $a$ and $b$, if the function $f$ is given by $f=-{\frac{r^{2}%
+a^{2}}{l^{2}}}$ (see below for precise definitions). We then show that $f$
must be of the type $f={\frac{n(r)}{\varrho^{2}}}$ for spacetime to be
circular. We give the transformations to the Schwarzschild or Boyer-Lindquist
coordinates which make these invariances manifest and show how, in $5D$, the
spacetimes are thus extended.\newline

Let us consider the four and five dimensional Kerr-Schild type metrics
\cite{KS} (for recent developments see Ref. \cite{Malek}),%
\begin{equation}
ds^{2}=d\bar{s}^{2}+f\,(\ell_{\mu}dx^{\mu})^{2} \label{eqn5}%
\end{equation}
where $d\bar{s}^{2}\equiv\bar{g}_{\mu\nu}dx^{\mu}dx^{\nu}$ is the line element
of four or five dimensional Minkowski spacetime in spheroidal coordinates
$x^{\mu}$, $d\bar{s}^{2}=-dt^{2}+ds_{(3)}^{2}$ or $d\bar{s}^{2}=-dt^{2}%
+ds_{(4)}^{2}$, see Eq.~(\ref{eqn2}) and Eq.~(\ref{eqn4}); where the four and
five dimensional vectors $\ell_{\mu}$ are null and geodesic and given by%

\begin{equation}%
\begin{array}
[c]{rcl}%
\ell_{\mu}dx^{\mu} & = & dt+{\frac{\varrho^{2}}{r^{2}+a^{2}}}dr+a\sin
^{2}\vartheta\,d\varphi\qquad\hbox{in 4D}\\[8pt]%
\ell_{\mu}dx^{\mu} & = & dt+\Delta_{r}dr+a\sin^{2}\vartheta\,d\varphi
+b\cos^{2}\vartheta\,d\psi\qquad\hbox{in 5D}\,,
\end{array}
\label{eqn6}%
\end{equation}
where $\varrho^{2}$ and $\Delta_{r}$ are defined in Eq.~(\ref{eqn2}) and
Eq.~(\ref{eqn4}). Note that it is not possible to set $x=r^{2}$ and let $x$ be
negative since $dr$ and hence $\ell_{\mu}\,dx^{\mu}$ and the metric
coefficients become complex.

As for $f$, it is for the moment an arbitrary function of $r$ and $\vartheta$.
(For the properties of the curvature tensors of such Kerr-Schild spacetimes
see, \textsl{e.g.}, Ref.~\cite{us09}.)\footnote{All that follows can easily be
extended to the case of an (anti-)de Sitter background (see Ref.~\cite{us09}%
).\smallskip}\newline

Obvious Killing vectors of these spacetimes are
\begin{equation}
\xi_{\mu}^{(k)}=\bar{\xi}_{\mu}^{(k)}+f\,\ell_{\mu}\ell_{k}\qquad
\text{with}\qquad\bar{\xi}_{\mu}^{(k)}\equiv\bar{g}_{\mu k} \label{eqn7}%
\end{equation}
where $k=t$, $k=\varphi$ and, in $5D$, $k=\psi$. It is easy to see that their
exterior products are linear in $f$~:
\begin{equation}
\xi^{(k_{1})}\wedge\xi^{(k_{2})}=\bar{\xi}^{(k_{1})}\wedge\bar{\xi}^{(k_{2}%
)}+f(\ell_{k_{2}}\bar{\xi}^{(k_{1})}-\ell_{k_{1}}\bar{\xi}^{(k_{2})}%
)\wedge\ell\label{eqn8}%
\end{equation}
and that their exterior derivatives are
\begin{equation}
\mathrm{d}\xi^{(k)}=\mathrm{d}\bar{\xi}^{(k)}+\mathrm{d}\left(  f\ell
_{k}\,\ell\right)  \,. \label{eqn9}%
\end{equation}

\textbf{Staticity: }

The condition of staticity is
\begin{equation}
\mathrm{d}\xi\wedge\xi=0 \label{eqn10}%
\end{equation}
where $\xi$ is any linear, timelike, combination of the Killing vectors. (See
\textsl{e.g.} Ref.~\cite{Eloy06} for clear definitions of staticity,
stationarity, axisymmetry and circularity.) As can easily be seen, the $4D$
Kerr-Schild spacetimes are static if and only if $\xi=\xi^{(t)}$, $a=0$ and
$f=f(r)$. As for the $5D$ Kerr-Schild spacetimes they are static if, either
\begin{equation}
a=b=0\quad,\quad\xi=\xi^{(t)}\quad\hbox{and}\quad f=f(r) \label{eqn11}%
\end{equation}
(in which case they are spherically symmetric), or
\begin{equation}
a^{2}=b^{2}\quad,\quad\xi=\xi^{(t)}+{\frac{b}{l^{2}-b^{2}}}(\xi^{(\varphi
)}+\xi^{(\psi)})\quad\hbox{and}\quad f=-{\frac{r^{2}+b^{2}}{l^{2}}}
\label{eqn12}%
\end{equation}
with $l^{2}$ a constant (not necessarily positive). The Killing vector $\xi$
has components
\begin{equation}
\xi_{\mu}=\left(  -{\frac{r^{2}+l^{2}}{l^{2}-b^{2}}},-{\frac{r^{2}}%
{l^{2}-b^{2}}},0,0,0\right)  \,. \label{eqn13}%
\end{equation}
Its norm is $\xi_{\mu}\xi^{\mu}=-{\frac{r^{2}+l^{2}}{l^{2}-b^{2}}}$. When
$l^{2}>0$ it is timelike if $l^{2}>b^{2}$. When $l^{2}<0$ it is timelike for
$r^{2}<-l^{2}$.

The metric of static spacetimes can be put in a form which is manifestly
time-independent and invariant under time-reversal:

$\bullet$ In the case $a=b=0$, see Eq.~(\ref{eqn11}), the transformation which
brings the metric into its Schwarzschild-like form reads%
\begin{equation}
dt=d\tau+{\frac{f(r)}{1-f(r)}}dr\quad\Longrightarrow\quad ds^{2}%
=-(1-f(r))d\tau^{2}+{\frac{dr^{2}}{1-f(r)}}\,+r^{2}d\Omega^{2} \label{eqn14}%
\end{equation}
where $d\Omega^{2}$ is the metric on a unit $2$ or $3$-sphere.

$\bullet$ In the $5D$ case when $a=b\neq0$, see Eq.~(\ref{eqn12}), the
transformation is given by%
\begin{equation}%
\begin{array}
[c]{rcl}%
dt & = & \sqrt{1-{\frac{b^{2}}{l^{2}}}}\,d\bar{t}-{\frac{\bar{r}\sqrt{\bar
{r}^{2}-b^{2}}}{\bar{r}^{2}+l^{2}-b^{2}}}\,d\bar{r}\quad,\quad r=\sqrt{\bar
{r}^{2}-b^{2}}\\[8pt]%
d\varphi & = & d\bar{\varphi}+{\frac{b\,l}{\sqrt{l^{2}-b^{2}}}}\,d\bar
{t}+{\frac{b(\bar{r}^{2}-b^{2})}{\bar{r}^{2}(\bar{r}^{2}+l^{2}-b^{2})}}%
\,d\bar{r}\\[8pt]%
d\psi & = & d\bar{\psi}+{\frac{b\,l}{\sqrt{l^{2}-b^{2}}}}\,d\bar{t}%
+{\frac{b(\bar{r}^{2}-b^{2})}{\bar{r}^{2}(\bar{r}^{2}+l^{2}-b^{2})}}\,d\bar{r}%
\end{array}
\label{eqn15}%
\end{equation}
which turns the metric into
\begin{equation}%
\begin{array}
[c]{rcl}%
ds^{2} & = & -{\frac{\bar{r}^{2}+l^{2}-b^{2}}{l^{2}}}\,d\bar{t}^{2}%
+{\frac{l^{2}}{\bar{r}^{2}+l^{2}-b^{2}}}\,d\bar{r}^{2}\\[8pt]
& + & \bar{r}^{2}\left(  d\vartheta^{2}+\sin^{2}\vartheta\,d\bar{\varphi}%
^{2}+\cos^{2}\vartheta\,d\bar{\psi}^{2}-{\frac{b^{2}}{l^{2}}}(\sin
^{2}\vartheta\,d\bar{\varphi}+\cos^{2}\vartheta\,d\bar{\psi})^{2}\right)
\end{array}
\label{eqn16}%
\end{equation}
so that its staticity is manifest. The coordinate transformation (\ref{eqn15})
requires $\bar{r}>b$, and $l^{2}>b^{2}$ if $l^{2}>0$. However, in view of the
form (\ref{eqn16}) of the metric, their range can be extended to all $l^{2}$
and to all $\bar{r}>0$. The surfaces of constant $\bar{r}$ are squashed
$3$-spheres, where $1-b^{2}/l^{2}$ parametrizes the squashing.\footnote{In
terms of the left-invariant forms of $SU(2)$ the metric of the squashed
$3$-sphere reads $d\Sigma_{_{\left(  3\right)  }}^{2}=\frac{1}{4}\left(
\sigma_{_{1}}^{2}+\sigma_{_{2}}^{2}+\left(  1-\frac{b^{2}}{l^{2}}\right)
\sigma_{_{3}}^{2}\right)  $.} When $b=0$ the metric is the one of (Anti-)de
Sitter spacetime, and for $b\neq0$ is asymptotically locally (A)dS, since the
curvature approaches a constant at infinity, i. e., $R_{\hspace{0.07in}%
\hspace{0.07in}\alpha\beta}^{\mu\nu}\rightarrow-l^{-2}\delta_{\alpha\beta
}^{\mu\nu}$.\newline

\textbf{Circularity:}

Let us now turn to the condition of circularity which guarantees that locally
the metric can be put in a form which is invariant under the simultaneous
inversion of time and angle(s)~:
\begin{equation}%
\begin{array}
[c]{rcl}
&  & \mathrm{d}\xi^{(k)}\wedge\xi^{(t)}\wedge\xi^{(\varphi)}=0\,,\qquad
\hbox{in 4D}\\[8pt]
&  & \mathrm{d}\xi^{(k)}\wedge\xi^{(t)}\wedge\xi^{(\varphi)}\wedge\xi^{(\psi
)}=0\qquad\hbox{in 5D}\,,
\end{array}
\label{eqn17}%
\end{equation}
for $k=t,$ $\varphi$ and, in $5D$, $k=t,\ \varphi,\ \psi$. In the case of
Kerr-Schild spacetimes, Eqs. (\ref{eqn17}) can be readily integrated, and one
finds that in $4D$ as well as in $5D$ these spacetimes are circular if and
only if
\begin{equation}
f={\frac{n(r)}{\varrho^{2}}}\,. \label{eqn18}%
\end{equation}

Now, the circularity property is closely connected to the existence of
Boyer-Lindquist coordinates \cite{BL}. Indeed, let us consider the
transformation defined as%
\begin{equation}
dt=dT+g(r)dr\ ,\ d\varphi=d\Phi+h_{\varphi}(r)dr\quad(\text{\textrm{and in}
}5D\mathrm{,}\ \text{ }d\psi=d\Psi+h_{\psi}(r)dr)\,. \label{eqn19}%
\end{equation}
As an easy calculation shows this coordinate transformation can eliminate the
cross terms in $dt\,dr$, $dr\,d\varphi$ (and, in $5D$, $dr\,d\psi$) if and
only if spacetime is circular, that is, if condition (\ref{eqn18}) is
satisfied. As for the functions $g$, $h_{\varphi}$ (and $h_{\psi}$) they are
given by
\begin{equation}
g(r)={\frac{c(r)}{\Delta}}\quad,\quad h_{\varphi}=-{\frac{a}{r^{2}+a^{2}}%
}{\frac{c(r)}{\Delta}}\quad,\quad\left(  h_{\psi}=-{\frac{b}{r^{2}+b^{2}}%
}{\frac{c(r)}{\Delta}}\right)  \label{eqn20}%
\end{equation}
where, recall, $f={\frac{n(r)}{\varrho^{2}}}$, and
\begin{equation}%
\begin{array}
[c]{rcl}%
c(r) & =n(r) & \quad\hbox{and}\quad\Delta=a^{2}+r^{2}-n(r)\qquad
\hbox{in 4D}\\[8pt]%
c(r) & =r^{2}n(r) & \quad\hbox{and}\quad\Delta=(a^{2}+r^{2})(b^{2}%
+r^{2})-r^{2}n(r)\qquad\hbox{in 5D}\,.
\end{array}
\label{eqn21}%
\end{equation}
Hence the $4D$ and $5D$ Kerr-Schild metrics, see Eqs.~(5),~(6), when spacetime
is circular, that is, fulfils condition (\ref{eqn17}) so that the
Boyer-Lindquist coordinates Eqs.~(19-21) exist, read:
\begin{equation}%
\begin{array}
[c]{rcl}%
ds_{(4)}^{2} & = & -dT^{2}+{\frac{\varrho^{2}}{\Delta}}dr^{2}+\varrho
^{2}d\vartheta^{2}+(a^{2}+r^{2})\sin^{2}\vartheta\,d\Phi^{2}+{\frac
{n(r)}{\varrho^{2}}}\left(  dT+a\sin^{2}\vartheta\,d\Phi\right)  ^{2}\\[8pt]%
ds_{(5)}^{2} & = & -dT^{2}+{\frac{r^{2}\varrho^{2}}{\Delta}}dr^{2}+\varrho
^{2}d\vartheta^{2}+(a^{2}+r^{2})\sin^{2}\vartheta\,d\Phi^{2}+(b^{2}+r^{2}%
)\cos^{2}\vartheta\,d\Psi^{2}\\[8pt]
&  & +{\frac{n(r)}{\varrho^{2}}}(dT+a\sin^{2}\vartheta\,d\Phi+b\cos
^{2}\vartheta\,d\Psi)^{2}\,.
\end{array}
\label{eqn22}%
\end{equation}
The invariance under the simultaneous inversion of time and angle(s) is thus
manifest in Boyer-Lindquist coordinates.

Note that, when $n(r)$ is an even function of $r$, if one sets $x=r^{2}$ and
let $x$ be negative, then the transformation from spheroidal to
Boyer-Lindquist coordinates Eqs.~(19-21) becomes complex, but the
Boyer-Lindquist metric coefficients, see Eq.~({\ref{eqn22}), remain real.}

Let us now explore whether this class of spacetimes admit event horizons.
Because of their isometries, an event horizon should generically be a null
surface $\Sigma$ of the form $r=r(\vartheta)$. In $5D$ the equation for this
class of null surfaces is given by%
\begin{equation}
\left(  {\frac{dr}{d\vartheta}}\right)  ^{2}=-\left[  {\frac{(r^{2}%
+a^{2})(r^{2}+b^{2})}{r^{2}}}-f\,\varrho^{2}\right]  \ . \label{NullHorizons}%
\end{equation}
Note that requiring circularity, fixes the form of the function $f$ as in
Eq.~(\ref{eqn18}), and hence the rhs of (\ref{NullHorizons}) depends only on
$r$, so that the equation that defines $\Sigma$ reduces to%
\begin{equation}
\left(  {\frac{dr}{d\vartheta}}\right)  ^{2}=-\frac{\Delta}{r^{2}}\ ,
\label{Nullsurfacesdelta}%
\end{equation}
where $\Delta$ is defined in Eq.~(\ref{eqn21}). Thus, since $\Delta>0$ for the
domain of outer communications, no such surfaces can exist in this region,
which is bounded by the surfaces defined by $\Delta=0$. If one assumes that
the function $n(r)$ is such that $\Delta$ admits a simple zero at $r=r_{+}%
=$constant, then $r=r_{+}$ is a null surface, which also turns out to be a
Killing horizon. Further null surfaces could exist in the region where
$\Delta<0$, but for them $r<r_{+}$, and they fail to be Killing horizons.

Since at the horizon the function $n\left(  r\right)  $ fulfills%
\[
n\left(  r_{+}\right)  =\frac{(r_{+}^{2}+a^{2})(r_{+}^{2}+b^{2})}{r_{+}^{2}%
}\ ,
\]
it is simple to compute the corresponding angular velocities, which remarkably
do not depend on the explicit form of $n\left(  r\right)  $. They are given by%
\begin{equation}
\Omega_{\phi} =\frac{a}{r_{+}^{2}+a^{2}}\ ,\quad,\quad\Omega_{\psi} =\frac
{b}{r_{+}^{2}+b^{2}}\ , \label{AngularVelocities}%
\end{equation}
and the surface gravity turns out to be%
\begin{equation}
\kappa=-\frac{1}{2}\frac{n^{\prime}\left(  r_{+}\right)  }{n\left(
r_{+}\right)  }+\frac{r_{+}^{4}-a^{2}b^{2}}{r_{+}\left(  r_{+}^{2}%
+a^{2}\right)  \left(  r_{+}^{2}+b^{2}\right)  }\ . \label{kappa}%
\end{equation}
The fact that neither the angular velocities nor the surface gravity depend on
$\vartheta$, reflects that rigidity of the event horizon is a consequence of
circularity, and does not require the use of field equations. In $4D$ this was
established by Carter \cite{Carter2}.

In sum, requiring circularity of Kerr-Schild spacetimes with flat seed,
implies the existence of Boyer-Lindquist coordinates, and also the rigidity of
event horizons, when they exist.

\section{Kerr-Schild metrics as solutions of Einstein and Einstein-Gauss-Bonnet
theories in vacuum}

Hitherto, no reference to field equations has been made. Let us now consider
the functions $f$ corresponding to the Kerr solution \cite{KS}, to the
Myers-Perry solution \cite{Myers86} (which solves the vacuum $5D$ Einstein
equations) and to the solution found in Ref.~\cite{us09} (which solves the
vacuum $5D$ Einstein-Gauss-Bonnet equations)\footnote{More precisely,
$f_{EGB}$ solves the field equations derived from the lagrangian
$R-\Lambda+{\frac{l^{2}}{4}}GB$ when their two maximally symmetric solutions
coincide, that is when $\Lambda=-{\frac{3}{l^{2}}}$. For $l^{2}>0$ the
maximally symmetric solution is anti-de Sitter spacetime; and for $l^{2}<0$ it
is de Sitter.\smallskip}, given by%
\begin{equation}
f_{K}={\frac{2m\,r}{\varrho^{2}}}\qquad,\qquad f_{MP}={\frac{2m}{\varrho^{2}}%
}\qquad,\qquad f_{EGB}=-{\frac{\varrho^{2}}{l^{2}}\ ,} \label{eqn23}%
\end{equation}
respectively, where $m$ and $l^{2}$ are constants.\footnote{Recall that
$\varrho^{2}=r^{2}+a^{2}\cos^{2}\vartheta$ or $\varrho^{2}=r^{2}+a^{2}\cos
^{2}\vartheta+b^{2}\sin^{2}\vartheta$ in $4D$ or $5D$, respectively.\smallskip
} Since $f_{K}$ and $f_{MP}$ are of the form $f={\frac{n(r)}{\varrho^{2}}}$,
we thus recover the well-known fact that the Kerr as well as the Myers-Perry
spacetimes are circular. The angular velocities and the surface gravity of the
horizons of the Myers-Perry solution are then obtained from
Eq.(\ref{AngularVelocities}), and Eq.(\ref{kappa}), with $n^{\prime}\left(
r_{+}\right)  =0$, respectively.

Note that the $5D$ Einstein-Gauss-Bonnet solution is generically \textsl{not}
circular, unless $b=a$, in which case it is static (see Eq.~\ref{eqn12}).
\newline

Let us now compare the Myers-Perry and Einstein-Gauss-Bonnet solutions. We
shall here limit ourselves to the evaluation of the curvature invariants.

It is usual (see, \textsl{e.g.}, Ref.~\cite{Gibbons04}) to study the
Myers-Perry black hole using Boyer-Lindquist coordinates. However, since the
rotating Einstein-Gauss-Bonnet spacetime is not circular no such coordinates
exist; thus we shall stick to the spheroidal coordinates\footnote{That we
shall henceforth call \textquotedblleft Kerr-Schild coordinates", as is
customary.\smallskip} in order to compare and contrast the two solutions. The
analysis becomes slightly more complicated since, as we saw in the previous
section, the Kerr-Schild coordinates cover a smaller portion of the
Myers-Perry spacetime than the Boyer-Lindquist ones (when $b\neq0$). \newline

The (asymptotically flat) Myers-Perry solution solves the vacuum $5D$ Einstein
equations, so the Kretschmann scalar is the simplest curvature invariant that
does not vanish:%
\begin{equation}
(R_{\mu\nu\rho\sigma}R^{\mu\nu\rho\sigma})_{MP}={\frac{96m^{2}}{\varrho^{12}}%
}(3\varrho^{4}-16\varrho^{2}r^{2}+16r^{4})\,. \label{eqn24}%
\end{equation}
As for the scalar curvature $R$ and the square of the Riemann tensor of the
(asymptotically (anti-)de Sitter) Einstein-Gauss-Bonnet solution, they are
given by%
\begin{equation}%
\begin{array}
[c]{rcl}%
R_{EGB} & = & -{\frac{4}{l^{2}\varrho^{2}}}(3\varrho^{2}+2r^{2})\ ,\\[8pt]%
(R_{\mu\nu\rho\sigma}R^{\mu\nu\rho\sigma})_{EGB} & = & {\frac{8}{l^{4}%
\varrho^{4}}}(9\varrho^{4}-12\varrho^{2}r^{2}+8r^{4})\,.
\end{array}
\label{eqn25}%
\end{equation}

Hence both solutions are manifestly singular if $\varrho^{2}\equiv r^{2}%
+a^{2}\cos^{2}\vartheta+b^{2}\sin^{2}\vartheta=0$ can vanish at $r\neq0$.
However, as we emphasized in the previous sections, $\varrho^{2}$ does
\textsl{not} vanish (if $b\neq0$) in the portion of spacetime covered by the
Kerr-Schild coordinates since the metric coefficients become complex for
$r^{2}<0$.

Now, in the case of the Myers-Perry solution, we know from the existence of
the Boyer-Lindquist coordinates that, in fact, $r^{2}$ \textsl{can} be
extended to negative values, so that $\varrho^{2}$ does vanish at
$r^{2}=-b^{2}$ and $\vartheta={\frac{\pi}{2}}$. This is the well-known
\textquotedblleft ring\textquotedblright\ singularity.

The special case $b=0$ must be considered separately for two reasons: first,
the Kerr-Schild and Boyer-Lindquist coordinates then cover the same portion of
spacetime, $r\in\lbrack0,\infty]$, so that the structure of the singularity
can be studied using Kerr-Schild coordinates only; and, second, since then,
$r=0$ when $\varrho^{2}=0$, the numerator of Eq.~(\ref{eqn24}) vanishes; but
$(R_{\mu\nu\rho\sigma}R^{\mu\nu\rho\sigma})_{MP}$ still diverges, though more
mildly, like $\varrho^{-8}$.\newline

In the case of the Einstein-Gauss-Bonnet solution on the other hand, only the
particular case $b=0$ can be studied using Kerr-Schild coordinates. As can be
seen from Eq (\ref{eqn25}) the curvature invariants are then \textsl{finite}
at $r=0$. Their values however depend on how we approach it. Indeed, the Ricci
scalar can take any value within the range $-{\frac{20}{l^{2}}}\leq
R_{EGB}\leq-{\frac{12}{l^{2}}}${.} For instance, in the \textquotedblleft
plane\textquotedblright\ $\vartheta={\frac{\pi}{2}}$: $R_{EGB}=-{\frac
{20}{l^{2}}}$ everywhere, including at $r=0$; but $R_{EGB}=-{\frac{12}{l^{2}}%
}$ at $r=0$ when we approach it from another fixed angle. Further values can
be obtained when approaching $r=0$ along given curves $\vartheta
=\vartheta\left(  r\right)  $ (a similar result holds for $(R_{\mu\nu
\rho\sigma}R^{\mu\nu\rho\sigma})_{EGB}$).

However, this apparent discontinuity at $r=0$ can be removed by an appropriate
choice of coordinates that suitably covers the region $r=0$. Indeed, if one
performs the following change of coordinates:
\begin{equation}
r=x\,\sin\,y\quad,\quad\cos\theta={\frac{x}{a}}\cos\,y
\end{equation}
which is valid (in a patch) around the origin, the Ricci and Kretschmann
scalars are given by%
\begin{align}
R_{EGB}  & =-{\frac{20}{\ell^{2}}}+{\frac{8}{\ell^{2}}}\cos^{2}y\ ,\\
(R_{\mu\nu\rho\sigma}R^{\mu\nu\rho\sigma})_{EGB}  & =\frac{40}{\ell^{4}}%
+\frac{32}{\ell^{4}}\cos^{2}y\cos2y\ ,
\end{align}
respectively, and the discontinuity is clearly removed. Note that within the
patch where the new coordinates are well defined, $R_{EGB}$ ranges as
expected, i.e. $-{\frac{20}{\ell^{2}}}\leq R_{EGB}\leq-{\frac{12}{\ell^{2}}}%
$.\newline

Let us finally mention the question of the existence of horizons in the $5D$
rotating solution of the Einstein-Gauss-Bonnet equations. Here again, only the
special case $b=0$ can be studied using Kerr-Schild coordinates. The equation
for the existence of null surfaces of the form $r=r\left(  \vartheta\right)
$, in (\ref{NullHorizons}), becomes
\begin{equation}
\left(  {\frac{dr}{d\vartheta}}\right)  ^{2}=-\left(  r^{2}+a^{2}%
+{\frac{\left(  r^{2}+a^{2}\cos^{2}\vartheta\right)  ^{2}}{l^{2}}}\right)  \,.
\end{equation}
In the asymptotically locally AdS case ($l^{2}>0$) this equation has
no solution because its rhs is manifestly negative.
We leave to further work the detailed analysis of the case  $l^2<0$.\\

As for the generic case $b\neq0$, the question of whether the solution is
regular everywhere and possesses horizons or not remains open.

\section*{Acknowledgments}

It is a pleasure to thank Julio Oliva, Misao Sasaki, and especially Eloy
Ayon-Beato for enlightening discussions. N. D. thanks the organizers and
sponsors of the workshop held in Valdivia at CECS in January 2010 and the
organizers and sponsors of the conference on \textquotedblleft Quantum Gravity
and the Foundations of Physics\textquotedblright\ celebrating Mario
Castagnino's 75th birthday held in Rosario in March 2010. A. A is an Alexander
von Humboldt fellow. R. T. wishes to thank the kind hospitality at the Yukawa
Institute for Theoretical Physics (YITP).

This research is partially funded by FONDECYT grants No 1085322, 1095098,
3110141, and by the Conicyt grant \textquotedblleft Southern Theoretical
Physics Laboratory\textquotedblright\ ACT-91. The Centro de Estudios
Cient\'{\i}ficos (CECS) is funded by the Chilean Government through the
Millennium Science Initiative and the Centers of Excellence Base Financing
Program of Conicyt. CECS is also supported by a group of private companies
which at present includes Antofagasta Minerals, Arauco, Empresas CMPC, Indura,
Naviera Ultragas and Telef\'{o}nica del Sur. CIN is funded by Conicyt and the
Gobierno Regional de Los R\'{\i}os.\appendix

\end{document}